\documentclass{emulateapj}

\usepackage{natbib}
\input{epsf.sty}

\shorttitle{...}
\shortauthors{HONG ET AL.}

\newcommand{\um}{$\mu$m}

\newcommand{\lstar}{$L^{*}$}
\newcommand{\lstarb}{$L_{B}^{*}$}
\newcommand{\lstaro}{$L_{3.6}^{*}$}
\newcommand{\lstarf}{$L_{8.0}^{*}$}

\newcommand{\mblimit}{$M_B>-18$}
\newcommand{\mblarge}{$M_B<-18$}
\newcommand{\lowmb}{$M_B>-16$}
\newcommand{\midmb}{$-18< M_B <-16$}

\newcommand{\solar}{\ifmmode_{\mathord\odot}\else$_{\mathord\odot}$\fi} 
\newcommand{\kms}{km\thinspace s$^{-1}$}     
\newcommand{\dgr}{\ifmmode {^\circ}\else $^\circ $\fi}  
\newcommand{\arcs}{\ifmmode {'' }\else $'' $\fi}  
\newcommand{\arcm}{\ifmmode {' }\else $' $\fi}    
\newcommand{\msolar}{\ifmmode {M_\odot} \else M$_\odot$\fi}
\newcommand{\zsun}{\ifmmode {Z_\odot} \else Z$_\odot$\fi}
\newcommand{\gapprox}{\ifmmode \buildrel > \over {_\sim} \else $\buildrel >\over {_\sim}$\fi}
\newcommand{\lapprox}{\ifmmode \buildrel < \over {_\sim} \else $\buildrel <\over {_\sim}$\fi}

\newcommand{\ha}{\ifmmode H\alpha \else H$\alpha$\fi}
\newcommand{\msunyr}{\ifmmode {M_\odot yr^{-1}} \else M$_\odot$ yr$^{-1}$\fi}
\newcommand{\darcs}{\ifmmode \buildrel {''} \over . \else \buildrel $''$ \over . \fi}

\begin{document}
\bibliographystyle{astron}

\title{Infrared Properties of a Complete Sample of Star-Forming Dwarf Galaxies}
\shorttitle{Infrared Properties of Star-Forming Dwarf Galaxies}

\author{Sukbum A. Hong\altaffilmark{1}, Jessica L. Rosenberg\altaffilmark{1}}
\affil{Department of Physics and Astronomy, George Mason University, Fairfax, VA
22030}
\email{shongg@gmu.edu, jrosenb4@gmu.edu}

\author{Matthew L. N. Ashby\altaffilmark{2}}
\affil{Harvard-Smithsonian Center for Astrophysics, 60 Garden Street, MS 65, Cambridge, MA 02138}
\email{mashby@cfa.harvard.edu}

\and

\author{John J. Salzer\altaffilmark{3}}
\affil{Astronomy Department, Indiana University, Bloomington, IN 47405}

\email{slaz@astro.indiana.edu}

\begin{abstract}

We present a study of a large, statistically complete sample of star-forming 
dwarf galaxies using mid-infrared observations from the {\it Spitzer Space Telescope}.
The relationships between metallicity,
star formation rate (SFR) and mid-infrared color in these systems show that
the galaxies span a wide range of properties. However, the galaxies do show a
deficit of 8.0 \um\ polycyclic aromatic hydrocarbon emission as is apparent from
the median 8.0 \um\  luminosity which is only 0.004 \lstarf\ while the median
$B$-band luminosity is 0.05 \lstarb. Despite many of the galaxies being 8.0 \um\ deficient, 
there is about a factor of 4 more extremely red galaxies in the [3.6] $-$ [8.0] color 
than for a sample of normal galaxies with similar optical colors. We show correlations 
between the [3.6] $-$ [8.0] color and luminosity, metallicity, and to a lesser extent
SFRs that were not evident in the original, smaller sample
studied previously. The luminosity--metallicity relation has a flatter slope
for dwarf galaxies as has been indicated by previous work. We also show a
relationship between the 8.0 \um\ luminosity and the metallicity of the galaxy
which is not expected given the competing effects (stellar mass, stellar
population age, and the hardness of the radiation field) that influence the 8.0
\um\ emission. This larger sample plus a well-defined selection function also
allows us to compute the 8.0 \um\ luminosity function and compare it with
the one for the local galaxy population. Our results show that below
10$^{9}$ $L$\solar, nearly all the 8.0 \um\ luminosity density of the local 
universe arises from dwarf galaxies that exhibit strong \ha\ emission -- i.e.,
8.0 \um\ and \ha\ selection identify similar galaxy populations despite the
deficit of 8.0 \um\ emission observed in these dwarfs.

\end{abstract}


\section{INTRODUCTION}

It has become clear that the mid-infrared properties of galaxies are dependent
on metallicity (e.g., \citealp{engelbracht2005, wu2006,rosenberg2006,
rosenberg2008, hunt2006,madden2006,ohalloran2006}). The
infrared emission, particularly from polycyclic aromatic hydrocarbon (PAH)
grains, is usually driven by the radiation from young stars
(\citealp{tielens2008} and references therein). However, the strength
of PAH emission is altered and may begin to break
down for lower metallicity systems. In the most extreme cases of SBS
0335-052 and I Zw 18, two of the lowest metallicity galaxies known, there is no
evidence for PAH emission \citep{houck2004, wu2007}. However these two sources
show dramatically different mid-infrared spectral energy distributions
(SEDs) that indicate the presence of a cold dust ($T$ \lapprox\ 30K)
component in I Zw 18 but not in SBS 0335-052. While metallicity
may influence the strength of the PAH features, these galaxies indicate that it
is not the key to understanding the mid-infrared slope in the lowest metallicity
systems. 

The correlation between the PAH emission from galaxies and metallicity has been
attributed to the harder radiation field in lower metallicity systems causing
the break-down of the PAH molecules \citep{madden2006,wu2006,ohalloran2006}
and to the delayed injection of PAH molecules into the interstellar medium due to
the youth of the stellar populations \citep{galliano2008}. Alternatively, the
shape of the mid-infrared continuum in galaxies, which is usually attributed to
the temperature of the dust and the star formation rate (SFR) in the galaxy, does not
seem to be dependent on the metallicity of the system. 

In this paper we continue the exploration of the relationship between the
mid-infrared properties of galaxies and their metallicity, luminosity, and
SFR. We build on the work presented by \citet{rosenberg2006}
for 19 galaxies by discussing the mid-infrared properties of much larger
samples of 86 galaxies presented in \S 3.1. We examine the relationship
between colors (\S 3.2), metallicity (\S 3.3), and SFR (\S 3.4) and 
the properties of galaxies. In \S 3.5, we use this expanded sample to
compute the 8.0 \um\ luminosity function for the star-forming dwarf galaxies
and compare it with that for local galaxies as a whole
\citep{huang2007}. We also use the 8.0 \um\ luminosity density as a measure 
of the star formation rate density (SFRD) and compare it with the value derived from 
the H$\alpha$ luminosity density.

Distances throughout this paper are based on $H_{0}$ = 70 km s$^{-1}$ Mpc$^{-1}$.
All magnitudes discussed in this paper are Vega relative magnitudes.

\section{OBSERVATIONS AND DATA REDUCTION}

The galaxies in this sample are selected from the KPNO International
Spectroscopic Survey (KISS, \citealp{salzer2000}), the first fully digital
objective prism survey for extragalactic emission-line objects. The galaxies
represent a star-forming dwarf (\mblimit) galaxy sub-sample of the KISS
galaxies selected by their \ha\ emission lines. The sample consists of all 60
KISS-selected dwarf galaxies in the 1.3$^\circ$ wide strip centered on
$\delta$(B1950)=29$^\circ$30$^\prime$ \citep{salzer2001} with velocities
less than 9000 \kms. We combine these data with 26 star-forming dwarf 
galaxies within the Bo\"{o}tes field where no velocity limit has been imposed,
16 of which have velocities greater than 9000 \kms. The Bo\"{o}tes field is a
8.5 deg$^{2}$ region that has been observed both as a part of KISS survey and
the NOAO Deep Wide-Field Survey (NDWFS; \citealp{jannuzi1999}).
There are 19 galaxies in this field that overlap with Spitzer Shallow Survey
area \citep{eisenhardt2004}. The inclusion of the higher velocity sample
increases the number of redder, more luminous sources but it does not change
the range of parameter space sampled as we show in the next section. The KISS
data for the full sample are presented in Table 1.

Observations of 67 star-forming dwarf galaxies at 3.6, 4.5, 5.8 and 8.0 \um\
were made in 2006 July using the Infrared Array Camera (IRAC; \citealp{fazio2004})
aboard the {\it Spitzer Space Telescope} \citep{werner2004}. The basic image
processing was reduced using the IRACproc package \citep{schuster2006}.
IRACproc is based on the {\it Spitzer} Science Center mosaic software MOPEX
(Mosaicking and Point-source extraction) and provides enhanced cosmic ray rejection.
The resulting mosaiced images are 130$''$ $\times$ 130$''$ except for KISSR 73
(259$''$$\times$ 259$''$) and KISSR 1048 (216$''$$\times$216$''$) because these
systems are more extended. The pixel scale for all images is 0\farcs86 per pixel.

\begin{figure}
\epsscale{0.9}
\plotone{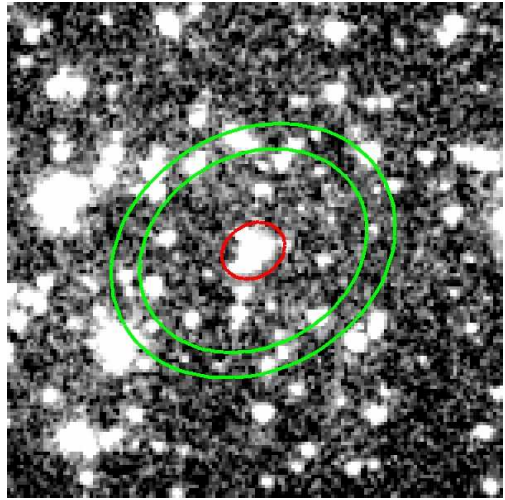}
\vspace{30pt}
\caption{3.6 \um\ image of KISSR 2378. The red (inner) ellipse shows the aperture in which
the galaxy flux was measured. The green (outer) ellipses show the annulus in which the sky
background was measured. This figure shows the degree to which stars were a problem
for the photometric measurements.}
\label{mopex}
\end{figure}

The measurement of photometry for these galaxies was divided into three processes:
point-source subtraction in the 3.6 and 4.5 \um\ bands, background subtraction,
and elliptical isophotal fitting. Stellar crowding was a problem for the galaxy
photometry and background estimation in the 3.6 and 4.5 \um\ bands as seen in the
3.6 \um\ image shown in Figure \ref{mopex}. The figure shows the aperture in which
we measure the galaxy flux (red) and the annulus we use for background subtraction
(green). MOPEX/APEX was used to identify and remove point sources from the mosaics,
except for sources found to be part of the galaxies in question. Some of the
bright point sources in the 3.6 and 4.5 \um\ bands leave residuals
after the subtraction. These residuals and any sources in the galaxy or background
area at 5.8 and 8.0 \um\ were masked before the isophotal fitting was performed
in those bands. After the point sources were subtracted and/or masked,
the background subtraction and isophotal fitting were done using the ELLIPSE
package within IRAF\footnote{IRAF is distributed by the National Optical Astronomy
Observatories, which is operated by the Association of Universities for Research in
Astronomy, Inc. (AURA) under cooperative agreement with the National Science
Foundation.}/STSDAS.  The ELLIPSE task was used to fit elliptical isophotes to
the galaxy given a fixed center, ellipticity, and position angle. The background was
defined as the average intensity in ellipses that are at least 5 pixels beyond the
extent of the galaxy (usually within an annulus at a radius of 36--45 pixels except
for the four largest systems for which larger annuli were used). The extent of the
galaxy was assumed to be the point at which the flux reaches within 1$\sigma$ of
the background level. For two of the galaxies (KISSR148 and KISSR 217)
bright point sources introduced a background gradient across the image. For these
sources, the IMSURFIT task within IRAF was used to fit a second-order polynomial
to the background in order to remove the gradient after the initial background
subtraction. After the image was background subtracted, the galaxy flux was
measured within the previously determined extent of the galaxy from a second run
of the ELLIPSE fitting routine.

The flux errors were assumed to be the
combination of the Poisson errors and a systematic error due to the uncertainty
in the galaxy extent. The flux error due to the uncertainty in the galaxies'
extent was computed to be the change in the measured flux when the galaxy extent
was increased or decreased by 1 pixel or by 10\%, whichever was larger. The results 
of the photometry are presented in Table 2. One galaxy, KISSR 85, was not detected 
in any of the IRAC bands and three additional sources, KISSR 73, KISSR 75, and 
KISSR 856, were only detected at 3.6 and 4.5 \um. We report upper limits for these
sources in Table 2. These limits for the galaxies that were detected at 3.6 \um,
were calculated as the number of pixels in the 3.6 \um\ band aperture multiplied
by the standard deviation of the background within the aperture at the appropriate
band. We calculated the upper limit for the galaxy that was not detected in any of
the bands as the standard deviation times the number of pixels within a 5 pixel
radius aperture.

\begin{table*}
\footnotesize
\caption{Properties of KISS Galaxies \label{tab:STATS}}
\begin{center}
\begin{tabular}{ccccccccccc}
\hline
KISSR   &    RA        &    Dec       &   $V_{hel}$   & $M_{B}$ & $M_{B_{0}}$\tablenotemark{a} & $(B - V)_{0}$\tablenotemark{b}
& c$_{H\beta}$\tablenotemark{c} & log $L_{H\alpha}$ & log[O/H]+12\tablenotemark{d} & log$(SFR)$\tablenotemark{e} \\
        &    J2000     &    J2000     & (km s$^{-1}$) &       &             &         &  & (erg s$^{-1}$) & & ($M_\odot$ yr$^{-1}$) \\
\hline
     1  &  12:15:06.9  &  29:01:10.0  &   7285  &  $-$ 17.73  &  $-$ 17.88  &   0.50  &    0.09  &  40.17  &   8.53  &  $-$ 0.87  \\
    32  &  12:21:11.4  &  29:32:08.7  &   7855  &  $-$ 16.61  &  $-$ 16.76  &   0.59  &    0.37  &  40.24  &   8.41  &  $-$ 0.59  \\
    40  &  12:22:23.6  &  29:26:37.8  &   7808  &  $-$ 17.83  &  $-$ 17.98  &   0.58  &    0.40  &  40.52  &   8.17  &  $-$ 0.28  \\
    49  &  12:24:35.2  &  29:27:32.0  &   7871  &  $-$ 17.42  &  $-$ 17.57  &   0.58  &    0.20  &  40.70  &   8.24  &  $-$ 0.25  \\
    52  &  12:25:25.6  &  29:44:18.2  &   7981  &  $-$ 18.00  &  $-$ 18.15  &   0.80  &    1.18  &  41.08  &     ...     &  $ $ 0.86  \\
    55  &  12:26:22.0  &  29:22:12.5  &   8067  &  $-$ 15.92  &  $-$ 16.07  &   0.19  &    0.45  &  40.14  &   7.84  &  $-$ 0.63  \\
    57  &  12:26:39.3  &  29:37:58.4  &   6595  &  $-$ 17.60  &  $-$ 17.75  &   0.38  &    0.64  &  40.38  &   8.49  &  $-$ 0.25  \\
    59  &  12:27:08.7  &  28:57:22.7  &   7447  &  $-$ 17.89  &  $-$ 18.04  &   0.65  &      ...     &  39.93  &     ...     &  $-$ 1.17  \\
    61  &  12:27:33.4  &  29:36:06.0  &   6560  &  $-$ 15.62  &  $-$ 15.77  &   0.52  &    0.09  &  39.78  &   8.10  &  $-$ 1.26  \\
    73  &  12:31:57.2  &  29:42:46.3  &     631  &  $-$ 14.16  &  $-$ 14.31  &   0.34  &    0.05  &  38.19  &   7.91  &  $-$ 2.87  \\
    75  &  12:32:48.2  &  29:23:27.3  &   7899  &  $-$ 17.26  &  $-$ 17.41  &   0.51  &    0.01  &  40.12  &     ...     &  $-$ 0.97  \\
    85  &  12:37:18.5  &  29:14:54.7  &   6939  &  $-$ 15.04  &  $-$ 15.19  &   0.11  &    0.00  &  39.88  &   7.60  &  $-$ 1.22  \\
    91  &  12:39:39.6  &  29:36:34.2  &   6989  &  $-$ 17.12  &  $-$ 17.27  &   0.52  &    0.34  &  40.48  &   8.49  &  $-$ 0.37  \\
  105  &  12:42:53.6  &  29:17:18.2  &   6717  &  $-$ 14.57  &  $-$ 14.72  &   0.48  &    0.01  &  39.62  &   7.80  &  $-$ 1.47  \\
  108  &  12:43:55.3  &  29:22:11.7  &   6982  &  $-$ 17.40  &  $-$ 17.55  &   0.36  &    0.24  &  40.54  &   8.23  &  $-$ 0.38  \\
  115  &  12:46:09.1  &  28:57:30.6  &   7019  &  $-$ 17.82  &  $-$ 17.97  &   0.59  &    0.31  &  40.46  &   8.47  &  $-$ 0.41  \\
  119  &  12:47:24.5  &  29:12:26.0  &   6916  &  $-$ 16.37  &  $-$ 16.52  &   0.50  &    0.74  &  40.35  &     ...     &  $-$ 0.20  \\
  125  &  12:48:38.4  &  29:11:24.9  &   7037  &  $-$ 17.76  &  $-$ 17.91  &   0.42  &    0.23  &  40.36  &   8.50  &  $-$ 0.57  \\
  133  &  12:51:06.6  &  29:11:48.2  &   6605  &  $-$ 16.97  &  $-$ 17.12  &   0.52  &    0.48  &  40.25  &   8.55  &  $-$ 0.50  \\
  142  &  12:53:49.2  &  28:56:33.1  &   7068  &  $-$ 17.77  &  $-$ 17.92  &   0.87  &      ...     &  39.79  &     ...     &  $-$ 1.31  \\
  148  &  12:54:45.3  &  28:55:29.8  &   2392  &  $-$ 15.53  &  $-$ 15.68  &   0.14  &    0.13  &  39.39  &   8.31  &  $-$ 1.62  \\
  156  &  12:57:43.6  &  29:00:11.9  &   6833  &  $-$ 16.95  &  $-$ 17.10  &   0.24  &    0.18  &  40.03  &   8.04  &  $-$ 0.94  \\
  170  &  13:00:31.2  &  28:57:01.5  &   6914  &  $-$ 17.19  &  $-$ 17.34  &   0.56  &    0.81  &  40.42  &   8.64  &  $-$ 0.08  \\
  171  &  13:00:37.2  &  28:39:50.7  &   7024  &  $-$ 16.99  &  $-$ 17.14  &   0.72  &    0.13  &  39.96  &   8.81  &  $-$ 1.05  \\
  182  &  13:02:25.7  &  28:51:28.9  &   6645  &  $-$ 17.63  &  $-$ 17.78  &   0.43  &    0.21  &  40.51  &   8.27  &  $-$ 0.44  \\
  187  &  13:04:16.8  &  28:51:02.5  &   7481  &  $-$ 16.26  &  $-$ 16.41  &   0.49  & $-$ 0.02  &  40.08  &   8.14  &  $-$ 1.04  \\
  191  &  13:04:38.8  &  28:58:21.8  &   7470  &  $-$ 17.95  &  $-$ 18.10  &   0.37  & $-$ 0.02  &  40.15  &   8.10  &  $-$ 0.96  \\
  192  &  13:05:06.5  &  28:38:28.5  &   5521  &  $-$ 15.86  &  $-$ 16.01  &   0.87  &    1.11  &  40.72  &     ...     &  $ $ 0.44  \\
  193  &  13:05:12.3  &  29:14:09.4  &   6971  &  $-$ 16.03  &  $-$ 16.18  &   0.50  &    0.85  &  40.12  &   8.23  &  $-$ 0.35  \\
  194  &  13:05:15.5  &  28:37:35.1  &   6555  &  $-$ 17.29  &  $-$ 17.44  &   0.49  &    0.68  &  40.45  &   8.33  &  $-$ 0.15  \\
  205  &  13:07:23.0  &  29:24:04.0  &   5211  &  $-$ 17.10  &  $-$ 17.25  &   0.38  &    0.39  &  40.24  &   8.06  &  $-$ 0.57  \\
  207  &  13:08:04.0  &  28:59:54.8  &   7332  &  $-$ 17.91  &  $-$ 18.06  &   0.45  &    1.99  &  41.56  &   8.06  &  $ $ 1.93  \\
  210  &  13:08:15.2  &  29:01:22.4  &   6131  &  $-$ 16.58  &  $-$ 16.73  &   0.34  &      ...     &  39.99  &     ...     &  $-$ 1.12  \\
  215  &  13:08:54.9  &  29:32:39.5  &   6764  &  $-$ 17.84  &  $-$ 17.99  &   0.59  &    1.11  &  40.82  &     ...     &  $ $ 0.54  \\
  217  &  13:09:07.3  &  28:40:06.6  &   5577  &  $-$ 17.59  &  $-$ 17.74  &   0.48  &    0.52  &  40.44  &   8.54  &  $-$ 0.27  \\
  236  &  13:13:35.3  &  29:07:35.2  &   5929  &  $-$ 17.38  &  $-$ 17.53  &   0.95  &      ...     &  39.80  &     ...     &  $-$ 1.30  \\
  238  &  13:14:37.6  &  29:19:04.6  &   6503  &  $-$ 17.30  &  $-$ 17.45  &   0.73  &    1.44  &  40.91  &     ...     &  $ $ 0.88  \\
  245  &  13:16:28.0  &  29:25:11.3  &   4754  &  $-$ 16.39  &  $-$ 16.54  &   0.48  &    0.49  &  39.96  &   8.39  &  $-$ 0.77  \\
  271  &  13:21:40.8  &  28:52:59.1  &   6737  &  $-$ 17.40  &  $-$ 17.55  &   0.67  &    0.60  &  40.54  &   8.31  &  $-$ 0.12  \\
  272  &  13:21:45.1  &  29:27:51.6  &   8912  &  $-$ 17.23  &  $-$ 17.38  &   0.34  &    0.01  &  39.99  &   7.91  &  $-$ 1.10  \\
  278  &  13:23:37.7  &  29:17:17.3  &   4062  &  $-$ 15.52  &  $-$ 15.67  &   0.40  &    0.03  &  39.30  &   8.02  &  $-$ 1.78  \\
  280  &  13:24:08.7  &  29:11:04.0  &   5463  &  $-$ 15.95  &  $-$ 16.10  &   0.52  &    0.02  &  39.56  &   8.06  &  $-$ 1.53  \\
  286  &  13:26:25.1  &  29:10:31.5  &   5161  &  $-$ 16.53  &  $-$ 16.68  &   0.54  &    0.22  &  40.73  &   8.06  &  $-$ 0.20  \\
  299  &  13:29:56.5  &  29:46:19.3  &   7779  &  $-$ 17.18  &  $-$ 17.33  &   0.25  &    0.28  &  40.46  &   8.24  &  $-$ 0.44  \\
  314  &  13:35:35.6  &  29:13:00.9  &     909  &  $-$ 15.27  &  $-$ 15.42  &   0.40  &    0.31  &  39.26  &   8.05  &  $-$ 1.61  \\
  386  &  13:54:25.8  &  29:33:00.2  &   7170  &  $-$ 17.69  &  $-$ 17.84  &   0.63  &    0.39  &  40.38  &   8.36  &  $-$ 0.43  \\
  396  &  13:57:10.0  &  29:13:10.1  &   2317  &  $-$ 15.03  &  $-$ 15.18  &   0.32  &      ...     &  39.63  &   7.91  &  $-$ 1.47  \\
  407  &  13:58:36.1  &  29:23:20.9  &   5726  &  $-$ 17.72  &  $-$ 17.87  &   0.47  &      ...     &  39.93  &     ...     &  $-$ 1.17  \\
  460  &  14:08:18.8  &  29:01:01.0  &   7295  &  $-$ 17.94  &  $-$ 18.09  &   0.84  &    0.64  &  40.20  &     ...     &  $-$ 0.42  \\
  505  &  14:16:55.4  &  29:29:11.5  &   3354  &  $-$ 16.60  &  $-$ 16.75  &   0.38  &    0.54  &  39.94  &   7.99  &  $-$ 0.76  \\
  507  &  14:17:24.7  &  29:41:11.8  &   8431  &  $-$ 17.67  &  $-$ 17.82  &   0.73  &      ...     &  40.08  &     ...     &  $-$ 1.02  \\
  541  &  14:23:58.3  &  29:49:33.8  &   3114  &  $-$ 16.33  &  $-$ 16.48  &   0.48  &    1.45  &  40.75  &     ...     &  $ $ 0.72  \\
  561  &  14:29:53.6  &  29:20:10.6  &   3683  &  $-$ 15.69  &  $-$ 15.83  &   0.29  &    0.02  &  39.74  &   8.08  &  $-$ 1.35  \\
  572  &  14:46:48.2  &  29:25:17.1  &   3794  &  $-$ 16.12  &  $-$ 16.27  &   0.39  & $-$ 0.08  &  39.56  &   8.34  &  $-$ 1.60  \\
  856  &  15:46:45.5  &  29:52:09.3  &   8385  &  $-$ 14.48  &  $-$ 14.63  &   0.63  &      ...     &  39.23  &     ...     &  $-$ 1.87  \\
  956  &  16:02:05.2  &  29:43:37.8  &   4410  &  $-$ 16.66  &  $-$ 16.81  &   0.40  &    0.69  &  40.15  &   8.33  &  $-$ 0.44  \\
1011  &  16:15:46.5  &  29:52:53.5  &   2580  &  $-$ 15.96  &  $-$ 16.11  &   0.52  &      ...     &  38.98  &     ...     &  $-$ 2.12  \\
1013  &  16:16:39.0  &  29:03:33.0  &   7540  &  $-$ 17.44  &  $-$ 17.59  &   0.37  &    0.03  &  40.14  &   8.07  &  $-$ 0.94  \\
1021  &  16:19:02.6  &  29:10:22.4  &   2708  &  $-$ 15.12  &  $-$ 15.27  &   0.27  &    0.13  &  39.40  &   8.03  &  $-$ 1.60  \\
1048  &  16:33:47.6  &  28:59:05.7  &   1100  &  $-$ 17.22  &  $-$ 17.37  &   0.37  &    0.02  &  38.01  &   8.32  &  $-$ 3.08  \\
2292  &  14:25:09.2  &  35:25:16.0  &   8659  &  $-$ 17.80  &  $-$ 18.31  &   0.66  &    0.28  &  40.39  &   8.41  &  $-$ 1.05  \\
2300  &  14:26:08.9  &  33:54:19.8  & 10271  &  $-$ 16.09  &  $-$ 16.09  &   0.50  &    0.15  &  40.52  &   7.89  &  $-$ 1.10  \\
2302  &  14:26:17.5  &  35:21:35.6  &   8342  &  $-$ 17.16  &  $-$ 17.16  &   0.42  &      ...     &  40.04  &     ...     &  $-$ 1.05  \\
2309  &  14:26:53.6  &  34:04:14.6  &   7231  &  $-$ 17.12  &  $-$ 17.12  &   0.43  &    0.28  &  40.09  &   7.98  &  $-$ 1.52  \\
2316  &  14:28:14.9  &  33:30:25.7  & 10685  &  $-$ 17.38  &  $-$ 18.57  &   0.81  &    1.09  &  41.01  &   9.10  &  $ $ 0.12  \\
\end{tabular}
\end{center}
\end{table*}

\begin{table*}
\footnotesize
\begin{center}
TABLE 1--Continued \\
\vspace*{10pt}
\begin{tabular}[t]{ccccccccccc}
\hline
\hline
KISSR   &    RA        &    Dec       &   $V_{hel}$   & $M_{B}$ & $M_{B_{0}}$\tablenotemark{a} & $(B - V)_{0}$\tablenotemark{b}
& c$_{H\beta}$\tablenotemark{c} & log $L_{H\alpha}$ & log[O/H]+12\tablenotemark{d} & log$(SFR)$\tablenotemark{e} \\
        &    J2000     &    J2000     & (km s$^{-1}$) &       &             &         &  & (erg s$^{-1}$) & & ($M_\odot$ yr$^{-1}$) \\
\hline
2318  &  14:28:24.6  &  35:10:21.5  & 22163  &  $-$ 17.88  &  $-$ 17.88  &   0.95  &      ...     &  41.15  &     ...     &  $ $ 0.05  \\
2322  &  14:29:09.6  &  32:51:26.8  &   8574  &  $-$ 17.84  &  $-$ 17.84  &   0.53  &    0.10  &  40.24  &   8.05  &  $-$ 1.30  \\
2326  &  14:29:32.7  &  33:30:40.4  &   7935  &  $-$ 17.14  &  $-$ 17.14  &   0.51  &    0.15  &  40.65  &   8.00  &  $-$ 0.96  \\
2338  &  14:30:27.9  &  35:32:07.2  & 11689  &  $-$ 17.20  &  $-$ 17.20  &   0.77  &    0.19  &  40.81  &   8.06  &  $-$ 0.74  \\
2344  &  14:31:03.6  &  35:31:14.8  &   4166  &  $-$ 17.75  &  $-$ 17.75  &   0.44  &    0.25  &  40.12  &   8.27  &  $-$ 1.40  \\
2346  &  14:31:14.4  &  33:19:13.2  & 10819  &  $-$ 16.80  &  $-$ 16.80  &   0.67  &    0.20  &  40.49  &   7.84  &  $-$ 1.15  \\
2349  &  14:31:20.0  &  34:38:03.8  &   4396  &  $-$ 16.63  &  $-$ 16.63  &   0.52  & $-$ 0.01  &  40.48  &   8.04  &  $-$ 1.10  \\
2357  &  14:31:39.2  &  33:26:32.3  & 10759  &  $-$ 17.67  &  $-$ 17.67  &   0.58  &    0.50  &  40.40  &   8.37  &  $-$ 1.05  \\
2359  &  14:31:49.3  &  35:28:40.0  & 22512  &  $-$ 17.55  &  $-$ 18.03  &   0.64  &    1.35  &  41.66  &   8.62  &  $ $ 0.32  \\
2368  &  14:32:18.9  &  33:02:53.7  & 10972  &  $-$ 17.05  &  $-$ 17.05  &   0.76  &    0.19  &  40.84  &   8.03  &  $-$ 0.74  \\
2378  &  14:33:19.4  &  32:43:00.0  & 11171  &  $-$ 17.79  &  $-$ 17.79  &   0.38  &    0.71  &  40.85  &   8.00  &  $ $ 0.27  \\
2382  &  14:34:08.0  &  34:19:34.5  &   6813  &  $-$ 17.81  &  $-$ 17.81  &   0.35  &    0.10  &  40.03  &   8.28  &  $-$ 1.40  \\
2384  &  14:34:30.1  &  32:50:46.1  & 11492  &  $-$ 17.72  &  $-$ 17.72  &   0.45  &    0.18  &  40.62  &   8.24  &  $-$ 0.35  \\
2398  &  14:36:33.1  &  34:58:04.5  &   9006  &  $-$ 18.01  &  $-$ 18.01  &   0.46  &    0.19  &  40.33  &   8.42  &  $-$ 1.10  \\
2400  &  14:37:04.8  &  33:00:20.4  &   3797  &  $-$ 16.54  &  $-$ 16.54  &   0.51  &    0.25  &  39.58  &   8.31  &  $-$ 1.34  \\
2403  &  14:37:42.6  &  33:36:26.7  & 12047  &  $-$ 16.24  &  $-$ 16.24  &   0.36  &    0.36  &  40.38  &   8.36  &  $-$ 1.05  \\
2406  &  14:38:27.8  &  35:08:59.1  &   8641  &  $-$ 17.89  &  $-$ 17.89  &   0.49  &    0.19  &  40.38  &   8.35  &  $-$ 1.05  \\
2407  &  14:38:29.3  &  33:20:07.5  & 20385  &  $-$ 17.51  &  $-$ 17.51  &   0.53  &    0.94  &  41.25  &     ...     &  $ $ 0.85  \\
2412  &  14:38:51.5  &  34:29:52.8  &   9845  &  $-$ 16.29  &  $-$ 16.29  &   0.62  &    0.15  &  40.45  &   7.95  &  $-$ 0.54  \\
2413  &  14:38:59.5  &  34:22:51.8  & 13110  &  $-$ 17.69  &  $-$ 19.04  &   0.86  &    0.61  &  40.77  &   9.03  &  $ $ 0.12  \\
2415  &  14:39:17.6  &  34:22:21.2  & 13265  &  $-$ 16.88  &  $-$ 17.96  &   0.76  &    1.14  &  41.07  &   8.71  &  $ $ 0.81  \\
\hline
\end{tabular}
\end{center}
\hspace*{10pt}$^{a}${ $B$-band absolute magnitude corrected for internal extinction using the ad hoc method from  \citep{melbourne2002}.} \\
\hspace*{10pt}$^{b}${ $B - V$ color corrected for Galactic absorption.} \\
\hspace*{10pt}$^{c}${ Decimal reddening coefficient.} \\
\hspace*{10pt}$^{f}${ A measure of metallicity. For detail of the calculation see \citet{melbourne2002} and \citet{salzer2005a}.} \\
\hspace*{10pt}$^{e}${ SFR calculated from the \ha\ luminosity using the prescription of \citet{kennicutt1998}.}
\end{table*}

\section{RESULTS}
\subsection{Properties of the Sample} \label{bozomath}

\begin{figure}
\epsscale{1.0}
\plotone{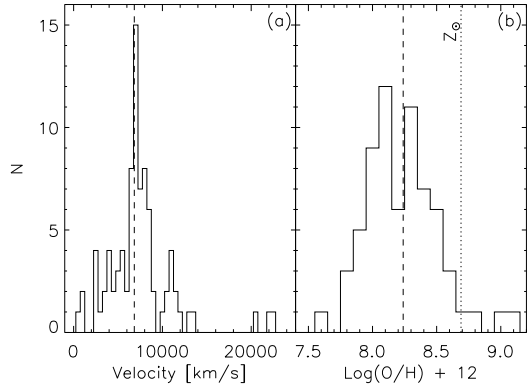}
\caption{Histogram of (a) velocity and (b) metallicity for the star-forming dwarf
galaxies. The dashed lines indicate the median values for the sample. The median
velocity is 6834 \kms, the median metallicity is log[O/H]+12 = 8.24
(0.35$Z_{\odot}$). The dotted line in panel (b) indicates solar metallicity,
log[O/H]+12 = 8.69 \citep{asplund2009}.}
\label{prope}
\end{figure}

The sample discussed here consists of two sub-samples of KISS galaxies: (1) 26
galaxies located in the Bo\"{o}tes field and (2) 60 galaxies in the KISS
30$^\circ$ strip. Figure \ref{prope}(a) shows the velocity distribution for the
sample. The median for the sample is 6834 \kms. Most of the galaxies have velocities
less than 9000 \kms, the limit imposed for systems in the  30$^\circ$ strip. In
discussing the properties of the galaxies we include 16 galaxies with velocities
greater than 9000 \kms\ that were observed in the Bo\"{o}tes field. Figure
\ref{prope}(b) shows the metallicity
distribution for 67 of the 86 galaxies. The remaining 19 galaxies do not have
strong enough lines for the metallicity determination. The median metallicity
for galaxies in the sample is log[O/H]+12 = 8.24 (0.35$Z$\solar) while the lowest
metallicity galaxy is only log[O/H]+12 = 7.60 (0.08$Z$\solar).

\begin{figure}
\plotone{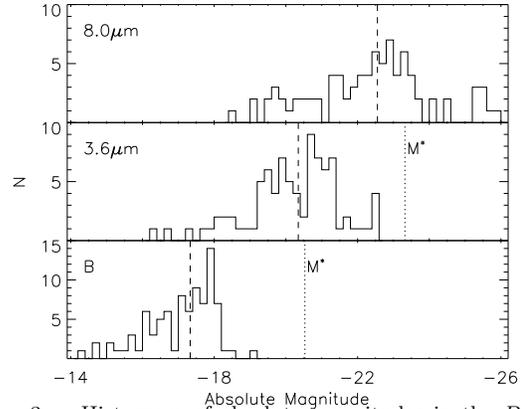}
\caption{Histogram of absolute magnitudes in the $B$, 3.6 $\mu$m and 8.0
$\mu$m bands. Magnitudes are relative to Vega. The dashed lines indicate
the median magnitudes in each band, $M_{B}$ = $-$17.33, $M_{3.6} $= $-$20.35,
and $M_{8.0}$ = $-$22.55. Note that these
galaxies have been selected to have $M_{B}$ $>$ $-$18 but ten galaxies are
brighter than that limit after being extinction corrected. The dotted lines
indicate the characteristic magnitudes, $M^{*}_{B}$ and $M^{*}_{3.6}$.
$M^{*}_{8.0}$ = $-$28.46 so it is off the right edge of the upper panel.}
\label{histomage}
\end{figure}

The distribution of $B$, 3.6 \um, and 8.0 \um\ absolute magnitudes is shown
in Figure \ref{histomage}. There are 10 galaxies brighter than
$M_{B_0}$ = $-$18 once they are corrected for extinction. The
median absolute magnitudes for the sample are $M_{B_0}$ = $-$17.33, $M_{3.6}$ =
$-$20.35, and $M_{8.0}$ = $-$22.55, which are indicated with dashed lines in the
figure. The $B$-band luminosities for the sample range from 0.003\lstarb\ to
0.25\lstarb\ \citep{driver2003} while the range is between 0.002\lstaro\ and
0.51\lstaro\ at 3.6 \um\ \citep{dai2009} and between 0.0001\lstarf\ and 0.09\lstarf\
at 8.0 \um\ \citep{huang2007}. The median 8.0 \um\ luminosity, 0.004\lstarf,
is a much smaller fraction of \lstar\ than the median $B$-band luminosity,
0.05\lstarb, or 3.6 \um\ luminosity, 0.06\lstaro. The smaller 8.0 \um\
luminosity with respect to \lstar\ indicates that, as a whole, these systems are
deficient in the 8.0 \um\ emission. This finding is consistent with a drop-off in
8.0 \um\ emission for low luminosity \citep{wu2009x} and low-metallicity
\citep{engelbracht2008} galaxies.

\begin{table*}
\footnotesize
\vspace*{10pt}
\caption{{\it Spitzer} IRAC Photometry \label{tab:STATS}}
\begin{center}
\begin{tabular}{ccccccccccc}
\hline
\hline
      &  $R_{3.6}$  &     &      &     &      &      &      &      &      &         \\
KISSR  & (arcsec)\tablenotemark{a} & Ellipticity\tablenotemark{b} & [3.6] & $\sigma_{3.6}$ & [4.5]
& $\sigma_{4.5}$ & [5.8] & $\sigma_{5.8}$ & [8.0] & $\sigma_{8.0}$   \\
\hline
    1  &  15.0  &  0.0  &   14.14  &   0.07  &   14.16  &   0.11  &   13.55  &   0.21  &   11.99  &   0.08  \\
   32  &   6.9  &  0.2  &   14.85  &   0.11  &   14.82  &   0.15  &   13.50  &   0.21  &   11.77  &   0.07  \\
   40  &  13.6  &  0.2  &   14.25  &   0.08  &   14.31  &   0.12  &   13.46  &   0.20  &   12.33  &   0.09  \\
   49  &   9.0  &  0.3  &   14.71  &   0.10  &   14.72  &   0.14  &   13.88  &   0.24  &   12.34  &   0.10  \\
   52  &  15.6  &  0.4  &   12.88  &   0.05  &   12.98  &   0.07  &   11.58  &   0.09  &    9.76  &   0.03  \\
   55  &   5.4  &  0.2  &   16.93  &   0.29  &   16.66  &   0.34  &   16.84  &   0.71  &   15.30  &   0.37  \\
   57  &  17.6  &  0.6  &   14.26  &   0.08  &   14.14  &   0.11  &   13.24  &   0.18  &   12.23  &   0.09  \\
   59  &  28.9  &  0.7  &   12.54  &   0.03  &   12.65  &   0.05  &   11.07  &   0.07  &    9.36  &   0.01  \\
   61  &   5.7  &  0.3  &   16.48  &   0.23  &   16.38  &   0.29  &   15.67  &   0.48  &   14.93  &   0.30  \\
   73  &  38.6  &  0.4  &   13.16  &   0.04  &   13.10  &   0.08  & $>$ 10.96 & \nodata & $>$ 10.34  & \nodata  \\
   75  &  12.6  &  0.0  &   15.34  &   0.06  &   15.41  &   0.08  & $>$ 13.76 & \nodata & $>$ 13.17 & \nodata  \\
   85  &   4.3  &  0.0  & $>$ 18.75 & \nodata & $>$ 17.79 & \nodata & $>$ 15.75 & \nodata & $>$ 15.04 & \nodata  \\
   91  &   9.6  &  0.2  &   14.03  &   0.08  &   14.03  &   0.11  &   12.50  &   0.14  &   10.81  &   0.05  \\
  105  &   4.3  &  0.0  &   17.64  &   0.39  &   17.32  &   0.46  &   16.59  &   0.74  &   15.27  &   0.41  \\
  108  &  12.3  &  0.6  &   14.75  &   0.10  &   14.71  &   0.14  &   13.95  &   0.25  &   12.37  &   0.09  \\
  115  &  10.5  &  0.0  &   13.79  &   0.07  &   13.79  &   0.09  &   13.13  &   0.18  &   11.56  &   0.06  \\
  119  &   8.9  &  0.3  &   15.54  &   0.15  &   15.67  &   0.22  &   15.26  &   0.42  &   13.65  &   0.18  \\
  125  &  10.1  &  0.3  &   13.82  &   0.07  &   13.83  &   0.10  &   12.51  &   0.14  &   10.88  &   0.05  \\
  133  &  15.8  &  0.6  &   13.94  &   0.07  &   14.01  &   0.11  &   12.84  &   0.16  &   11.20  &   0.05  \\
  142  &  21.2  &  0.4  &   13.03  &   0.05  &   13.21  &   0.08  &   12.45  &   0.13  &   11.66  &   0.06  \\
  148  &  18.4  &  0.4  &   14.17  &   0.07  &   14.06  &   0.09  &   13.15  &   0.18  &   12.03  &   0.06  \\
  156  &  12.5  &  0.0  &   14.99  &   0.12  &   15.02  &   0.16  &   14.65  &   0.34  &   13.28  &   0.14  \\
  170  &  25.8  &  0.7  &   13.86  &   0.07  &   14.05  &   0.11  &   12.83  &   0.15  &   11.26  &   0.06  \\
  171  &  11.8  &  0.4  &   14.34  &   0.09  &   14.31  &   0.12  &   13.51  &   0.21  &   12.83  &   0.06  \\
  182  &  10.3  &  0.0  &   14.27  &   0.08  &   14.37  &   0.12  &   13.48  &   0.21  &   12.06  &   0.08  \\
  187  &  10.0  &  0.5  &   15.83  &   0.18  &   15.73  &   0.23  &   15.10  &   0.44  &   13.57  &   0.15  \\
  191  &  12.2  &  0.0  &   14.11  &   0.08  &   14.09  &   0.11  &   13.46  &   0.20  &   12.15  &   0.09  \\
  192  &  18.8  &  0.7  &   14.77  &   0.10  &   14.71  &   0.14  &   13.75  &   0.23  &   12.23  &   0.09  \\
  193  &   6.0  &  0.3  &   16.23  &   0.20  &   16.24  &   0.28  &   15.72  &   0.50  &   14.63  &   0.25  \\
  194  &   8.2  &  0.3  &   14.28  &   0.08  &   14.31  &   0.12  &   13.54  &   0.20  &   12.19  &   0.09  \\
  205  &  10.2  &  0.0  &   14.46  &   0.09  &   14.40  &   0.12  &   14.06  &   0.26  &   12.81  &   0.11  \\
  207  &  25.0  &  0.8  &   14.12  &   0.08  &   14.26  &   0.12  &   13.53  &   0.21  &   12.02  &   0.07  \\
  210  &  10.2  &  0.1  &   15.15  &   0.12  &   15.09  &   0.16  &   14.48  &   0.31  &   12.90  &   0.12  \\
  215  &  13.1  &  0.4  &   13.21  &   0.05  &   13.16  &   0.06  &   12.08  &   0.11  &   10.40  &   0.04  \\
  217  &  14.4  &  0.0  &   13.65  &   0.06  &   13.59  &   0.07  &   12.53  &   0.13  &   11.12  &   0.06  \\
  236  &  20.4  &  0.4  &   12.13  &   0.03  &   12.21  &   0.05  &   11.00  &   0.07  &    9.23  &   0.02  \\
  238  &  11.3  &  0.3  &   13.68  &   0.07  &   13.73  &   0.09  &   13.47  &   0.20  &   12.71  &   0.11  \\
  245  &   8.7  &  0.0  &   14.48  &   0.09  &   14.54  &   0.14  &   13.32  &   0.19  &   11.83  &   0.07  \\
  271  &   9.5  &  0.0  &   14.04  &   0.08  &   13.93  &   0.10  &   13.35  &   0.19  &   11.96  &   0.08  \\
  272  &  10.8  &  0.1  &   14.81  &   0.11  &   14.68  &   0.15  &   14.52  &   0.29  &   13.22  &   0.14  \\
  278  &   9.2  &  0.2  &   15.39  &   0.14  &   15.33  &   0.19  &   15.38  &   0.46  &   14.33  &   0.25  \\
  280  &   9.8  &  0.4  &   15.47  &   0.15  &   15.53  &   0.21  &   15.77  &   0.55  &   14.69  &   0.26  \\
  286  &   7.6  &  0.0  &   14.35  &   0.09  &   14.24  &   0.12  &   13.25  &   0.19  &   11.78  &   0.08  \\
  299  &   5.9  &  0.0  &   15.05  &   0.12  &   14.94  &   0.16  &   14.14  &   0.27  &   12.59  &   0.10  \\
  314  &  32.6  &  0.4  &   12.46  &   0.03  &   12.80  &   0.05  &   12.97  &   0.12  &   11.39  &   0.04  \\
  386  &  15.5  &  0.6  &   14.04  &   0.08  &   14.07  &   0.11  &   13.36  &   0.20  &   11.92  &   0.08  \\
  396  &   9.2  &  0.0  &   14.93  &   0.11  &   14.78  &   0.14  &   14.50  &   0.32  &   14.02  &   0.20  \\
  407  &  15.1  &  0.2  &   13.81  &   0.06  &   13.81  &   0.09  &   13.19  &   0.18  &   11.82  &   0.08  \\
  460  &  22.0  &  0.5  &   12.53  &   0.04  &   12.59  &   0.06  &   11.28  &   0.08  &    9.58  &   0.03  \\
  505  &  14.2  &  0.4  &   14.00  &   0.07  &   13.94  &   0.10  &   13.55  &   0.21  &   12.81  &   0.10  \\
  507  &  12.6  &  0.3  &   14.18  &   0.08  &   14.19  &   0.11  &   13.66  &   0.22  &   12.14  &   0.09  \\
  541  &  14.5  &  0.4  &   13.70  &   0.06  &   13.66  &   0.09  &   13.15  &   0.18  &   12.39  &   0.08  \\
  561  &  12.4  &  0.4  &   15.44  &   0.14  &   15.43  &   0.19  &   15.41  &   0.42  &   14.57  &   0.25  \\
  572  &  16.7  &  0.6  &   14.43  &   0.09  &   14.37  &   0.12  &   13.46  &   0.20  &   12.28  &   0.09  \\
  856  &   5.2  &  0.0  &   15.33  &   0.14  &   15.23  &   0.18  & $>$ 15.39 & \nodata & $>$ 14.80 & \nodata  \\
  956  &  13.5  &  0.5  &   14.11  &   0.08  &   14.11  &   0.11  &   13.55  &   0.21  &   12.17  &   0.09  \\
 1011  &  23.8  &  0.5  &   13.66  &   0.06  &   13.86  &   0.10  &   13.58  &   0.21  &   12.23  &   0.09  \\
 1013  &  11.8  &  0.6  &   15.21  &   0.13  &   15.15  &   0.17  &   14.84  &   0.36  &   13.56  &   0.15  \\
 1021  &   9.0  &  0.1  &   15.09  &   0.12  &   15.04  &   0.16  &   14.76  &   0.36  &   13.25  &   0.15  \\
 1048  &  62.3  &  0.3  &   10.82  &   0.01  &   10.59  &   0.01  &    9.59  &   0.02  &    8.49  &   0.01  \\
 2292  &  12.0  &  0.1  &   14.31  &   0.07  &   14.29  &   0.09  &   13.53  &   0.17  &   12.15  &   0.07  \\
 2300  &  10.0  &  0.3  &   16.59  &   0.19  &   16.32  &   0.24  &   15.96  &   0.51  &   14.40  &   0.19  \\
 2302  &  14.0  &  0.3  &   16.05  &   0.15  &   16.03  &   0.21  &   15.81  &   0.47  &   15.04  &   0.25  \\
 2309  &  12.0  &  0.2  &   15.40  &   0.11  &   15.43  &   0.16  &   14.92  &   0.32  &   14.18  &   0.17  \\
 2316  &  14.0  &  0.1  &   14.15  &   0.06  &   14.09  &   0.08  &   12.61  &   0.11  &   10.61  &   0.03  \\

\end{tabular}
\end{center}
\footnotesize
\end{table*}

\begin{table*}\setcounter{table}{2}
\footnotesize
\begin{center}
TABLE 2--Continued \\
\vspace*{10pt}
\begin{tabular}{ccccccccccc}
\hline
      &  $R_{3.6}$  &     &      &     &      &      &      &      &      &         \\
KISSR  & (arcsec)\tablenotemark{a} & Ellipticity\tablenotemark{b} & [3.6] & $\sigma_{3.6}$ & [4.5]
& $\sigma_{4.5}$ & [5.8] & $\sigma_{5.8}$ & [8.0] & $\sigma_{8.0}$   \\
\hline
 2318  &  14.0  &  0.3  &   15.50  &   0.11  &   15.31  &   0.15  &   14.68  &   0.28  &   12.27  &   0.07  \\
 2322  &  20.0  &  0.2  &   14.76  &   0.08  &   14.68  &   0.11  &   14.31  &   0.24  &   12.93  &   0.10  \\
 2326  &  14.0  &  0.2  &   15.39  &   0.11  &   15.33  &   0.15  &   14.56  &   0.27  &   13.18  &   0.11  \\
 2338  &  14.0  &  0.2  &   15.82  &   0.13  &   15.49  &   0.16  &   14.60  &   0.27  &   12.89  &   0.09  \\
 2344  &  70.0  &  0.7  &   12.66  &   0.03  &   12.53  &   0.04  &   11.56  &   0.07  &   10.80  &   0.04  \\
 2346  &  12.0  &  0.1  &   15.88  &   0.14  &   15.80  &   0.19  &   15.25  &   0.37  &   13.67  &   0.14  \\
 2349  &  20.0  &  0.3  &   14.39  &   0.07  &   14.10  &   0.09  &   12.94  &   0.13  &   11.47  &   0.05  \\
 2357  &  20.0  &  0.6  &   15.19  &   0.10  &   15.01  &   0.13  &   14.89  &   0.31  &   13.42  &   0.12  \\
 2359  &  12.0  &  0.1  &   16.14  &   0.16  &   15.96  &   0.20  &   15.69  &   0.45  &   13.03  &   0.10  \\
 2368  &  14.0  &  0.4  &   15.89  &   0.14  &   15.45  &   0.16  &   14.57  &   0.27  &   12.75  &   0.09  \\
 2378  &   8.5  &  0.2  &   15.67  &   0.15  &   15.42  &   0.18  &   15.07  &   0.40  &   14.18  &   0.21  \\
 2382  &  20.0  &  0.4  &   14.17  &   0.06  &   14.16  &   0.09  &   13.47  &   0.16  &   12.07  &   0.06  \\
 2384  &   6.9  &  0.2  &   15.43  &   0.14  &   15.28  &   0.18  &   14.60  &   0.32  &   13.11  &   0.13  \\
 2398  &  30.0  &  0.4  &   14.27  &   0.07  &   14.17  &   0.09  &   13.39  &   0.16  &   11.98  &   0.06  \\
 2400  &  24.0  &  0.7  &   13.96  &   0.07  &   14.08  &   0.11  &   13.40  &   0.19  &   12.40  &   0.09  \\
 2403  &  14.0  &  0.3  &   16.96  &   0.23  &   16.78  &   0.29  &   15.81  &   0.47  &   14.88  &   0.23  \\
 2406  &  20.0  &  0.1  &   14.25  &   0.06  &   14.26  &   0.09  &   13.40  &   0.16  &   11.93  &   0.06  \\
 2407  &   8.6  &  0.6  &   16.03  &   0.19  &   15.88  &   0.24  &   15.58  &   0.47  &   14.72  &   0.28  \\
 2412  &   5.6  &  0.0  &   16.28  &   0.21  &   16.21  &   0.27  &   16.17  &   0.59  &   14.33  &   0.24  \\
 2413  &   8.2  &  0.0  &   13.99  &   0.08  &   13.93  &   0.10  &   12.63  &   0.14  &   10.52  &   0.04  \\
 2415  &   5.9  &  0.0  &   14.87  &   0.11  &   14.72  &   0.14  &   13.26  &   0.19  &   11.15  &   0.05  \\
 \hline

\end{tabular}
\end{center}
\hspace*{80pt}$^{a}${ The semimajor axis of the aperture in which the flux was measured.} \\
\hspace*{80pt}$^{b}${ The ellipticity of the fit aperture.}
\footnotesize
\end{table*}

\subsection{Colors of Star-forming Dwarf Galaxies } \label{bozomath}

\begin{figure}
\plotone{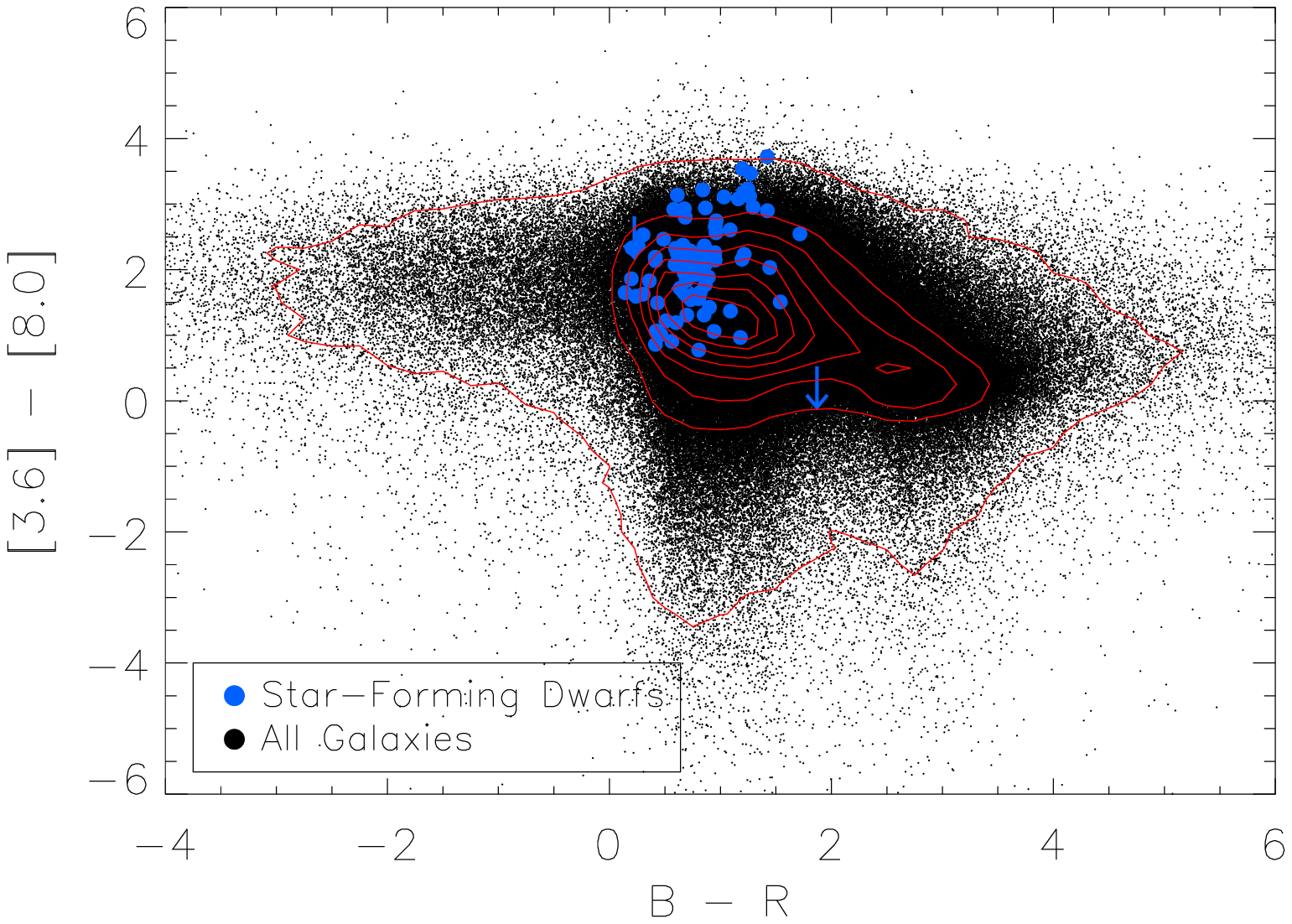}
\caption{[3.6] $-$ [8.0] vs. $B - R$ colors for the star-forming dwarf galaxies
(blue circles) and for galaxies in both the SDWFS \citep{ashby2009, brown2007} and
NDWFS \citep{jannuzi1999} with $m_{3.6}$ $<$ 20 (black circles). Two galaxies 
have upper limits at the [3.6] $-$ [8.0] are indicated by arrows. Contours
indicate the density of the black circles.}
\label{shallowe}
\end{figure}

Figure \ref{shallowe} shows the relationship between the [3.6] $-$ [8.0] and $B - R$ colors for
galaxies in this sample (large blue circles) and for all sources
classified as galaxies (black circles) in both the Spitzer Deep, Wide-Field Survey
(SDWFS; \citealp{ashby2009, brown2007}) and the NDWFS
\citep{jannuzi1999}. The $R$-band magnitudes are derived from the Sloan Digital
Sky Survey \citep{adelman-mcCarthy2008} $r$-band magnitudes transformed to
Johnson $R$ using the color correction from \citet{fadda2004}. Figure \ref{shallowe}
shows that the star-forming dwarf galaxies are significantly bluer in their optical
colors than the majority of galaxies because their light is dominated by star
formation. The [3.6] $-$ [8.0] color spans the same range covered by
late-type galaxy colors in the SDWFS but tend toward the redder end of the
galaxy distribution including a few extremely red sources. The star-forming dwarf galaxies
with $0 <$ $B - R$ $< 2$ and [3.6] $-$ [8.0] $> 2.5$ represent almost 40\% of the sample,
approximately ten times the fraction of SDWFS galaxies in this range and almost four times
the fraction of SDWFS galaxies in this range when only galaxies with
$0 <$ $B - R$ $< 2$ colors are considered. Expanding this analysis beyond the 19-galaxy
sample of \citet[hereafter Paper I]{rosenberg2006} to the 86 galaxies treated here has
helped to fill in this color-color plot and improve the statistics, but it has not
significantly changed the range of galaxy colors identified except to pick up a
few galaxies with redder $B - R$ colors.

\begin{figure}
\plotone{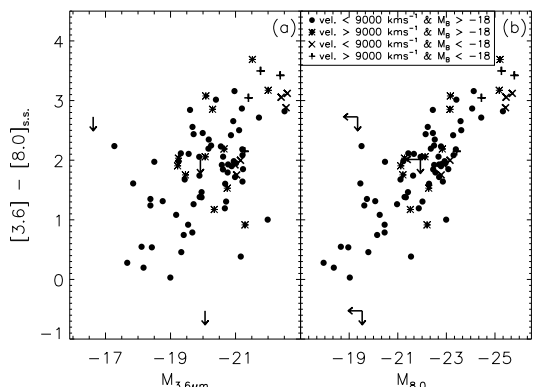}
\caption{[3.6] $-$ [8.0]$_{S.S.}$ color as a function of the 3.6 and stellar
subtracted 8.0 \um\ absolute magnitude. The filled circles indicate galaxies with
velocities smaller than 9000 \kms\ and with absolute $B$ magnitudes dimmer than $-$18.
The asterisks indicate galaxies with velocities larger than 9000 \kms\ and with absolute
$B$ magnitudes dimmer than $-$18. The crosses indicate galaxies with velocities smaller
than 9000 \kms\ and with absolute $B$ magnitudes brighter than $-$18. The plus signs
indicate galaxies with velocities larger than 9000 \kms\ and with absolute $B$
magnitudes brighter than $-$18. Several of the galaxies have upper limits only at
8.0 \um\ and are indicated by arrows in both color and, in the $M_{8.0}$ plot,
absolute magnitude.}
\label{midc4}
\end{figure}

Figure \ref{midc4} presents the [3.6] $-$ [8.0]$_{S.S.}$ color as a function of the
3.6 and stellar subtracted 8.0 \um\ luminosities. The 8.0 \um\ magnitude and luminosity
has had subtracted using a scale factor of 0.227 of 3.6 \um\ \citep{pahre2004a} as a
function of the 3.6 and stellar subtracted 8.0 \um\ luminosities. The stellar subtracted
8.0 \um\ flux used here and in some of the figures below as a better measure of the dust
emission in this band. The stellar subtracted magnitude is not used in the color-color plots
because galaxies with purely stellar colors can not be plotted for comparison since 8.0 \um\
flux drops to zero. In general, the more luminous galaxies are redder in the
[3.6] $-$ [8.0]$_{S.S.}$ color, a trend that was not clear for the 19 galaxies in
Paper I. Since the [3.6] $-$ [8.0]$_{S.S.}$ color is a good proxy for
the mass-normalized dust content (8.0 \um\ probes PAH plus hot dust continuum
while 3.6 \um\ traces the stellar mass), redder galaxies being more luminous
indicates that the dust emission per unit stellar mass increases both as
stellar mass (3.6 \um\ luminosity) and total dust emission from the galaxy (8.0
\um\ luminosity) increase. The mass normalized dust content of a galaxy appears
to be better correlated with the total dust content of the system (correlation
coefficient $= -0.79$) than it is with the total stellar mass (correlation
coefficient $= -0.45$). This difference in correlation seems to indicate that
the process that regulates the average dust emission in these galaxies --
either the delayed injection of dust due to the youth of the stellar population
or the destruction of dust in the hard stellar radiation field -- has a
significant impact on the total dust luminosity of the galaxy but is not
strongly tied to the bulk stellar mass.

\begin{figure}
\plotone{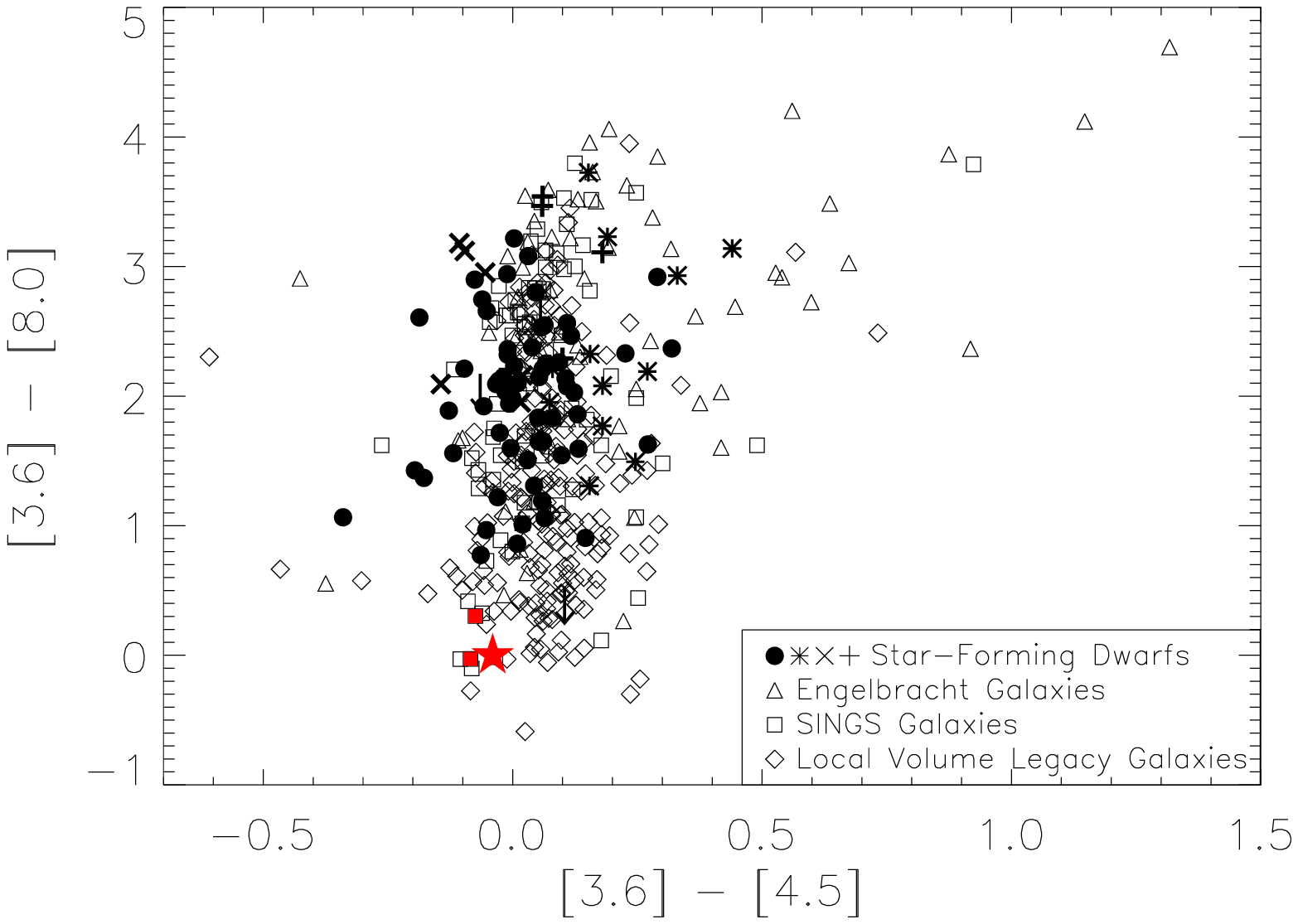}
\caption{[3.6] $-$ [8.0] vs. [3.6] $-$ [4.5] color--color diagram. The galaxies in
our sample are shown by the same symbols as in Figure \ref{midc4}. 
Star-forming galaxies from \citet{engelbracht2005} are shown as open triangles, SINGS
\citep{kennicutt2003} are shown as open squares, and LVL galaxies
\citep{dale2009} are shown as open diamonds. The red star indicates the theoretical
colors of M0 {III} stars \citep{pahre2004b} for reference. In addition, the red filled
squares indicate the observed colors of elliptical galaxies with the dust-free fitting
SED model in SINGS.}
\label{compu}
\end{figure}

Figure \ref{compu} shows the [3.6] $-$ [8.0] versus [3.6] $-$ [4.5] color--color plot
for the galaxies in this sample, starburst galaxies from \citet{engelbracht2005}, galaxies
from {\it Spitzer} Infrared Nearby Galaxies Survey (SINGS; \citealp{kennicutt2003}), and galaxies
from the {\it Spitzer} Local Volume Legacy (LVL; \citealp{dale2009}). The SINGS samples
cover a wide range of morphologies, metallicities, SFRs, and infrared-to-optical
ratios in normal galaxies \citep{kennicutt2003}. The LVL sample includes all galaxies within
11 Mpc down to a magnitude limit of $m_{B}$ $=$ 15. This complete sampling of the local
volume includes the full range of galaxy types from ellipticals to spirals and irregulars.
The [3.6] $-$ [8.0] color of the KISS galaxies spans the blue end of the range covered by
the LVL galaxies. Our sample does not have galaxies as red in the [3.6] $-$ [8.0] color as
those in LVL because we do not include early-type systems. There is little correlation between
the [3.6] $-$ [8.0] and [3.6] $-$ [4.5] colors with the exception of the reddest galaxies
([3.6] $-$ [4.5] $>0.4$) which are all red in the [3.6] $-$ [8.0] color ([3.6] $-$ [8.0] $>$ 1.5).
Most of the galaxies fall within a relatively small range of the [3.6] $-$ [4.5] color around
the colors of a theoretical M0 III star \citep{pahre2004b} and 
the SINGS elliptical galaxies NGC 0584 and NGC 1404 \citep{dale2007}. We use NGC 0584 and 1404 as fiducial
dust-free ellipticals because their spectra show no evidence for dust emission in
the mid-infrared\footnote{See http://web.ipac.caltech.edu/staff/jarrett/rac/carlibration\\
\hspace*{15pt}/galaxies.html.}.
The galaxies with the reddest [3.6] $-$ [4.5] color ([3.6] $-$ [4.5] $>$ 0.7) are
primarily luminous systems from \citet{engelbracht2005}. The lack of dwarf galaxies
with these red colors indicates that hot dust does not make a significant contribution
to the 4.5 \um\ emission in these galaxies. The [3.6] $-$ [4.5] colors are bluer than
the theoretical M0 III stellar color and the fidical dust-free elliptical galaxies
for a few of the galaxies possibly indicating emission in the 3.6 \um\ band from the 3.3 \um\
PAH feature.

\subsection{Effect of Metallicity on the Properties of Star-forming Dwarf Galaxies} \label{bozomath}

The luminosity--metallicity relationship is fundamental to galaxies and their
evolution, connecting the star formation history in galaxies to the increase in
metal abundance. \citet{tremonti2004} used the SDSS to
extend this relationship to a stellar mass--metallicity relationship and found that
the lower mass systems are more prone to metal loss, probably through galaxy
winds that preferentially remove metals from the lower potential systems. 

\begin{figure}
\plotone{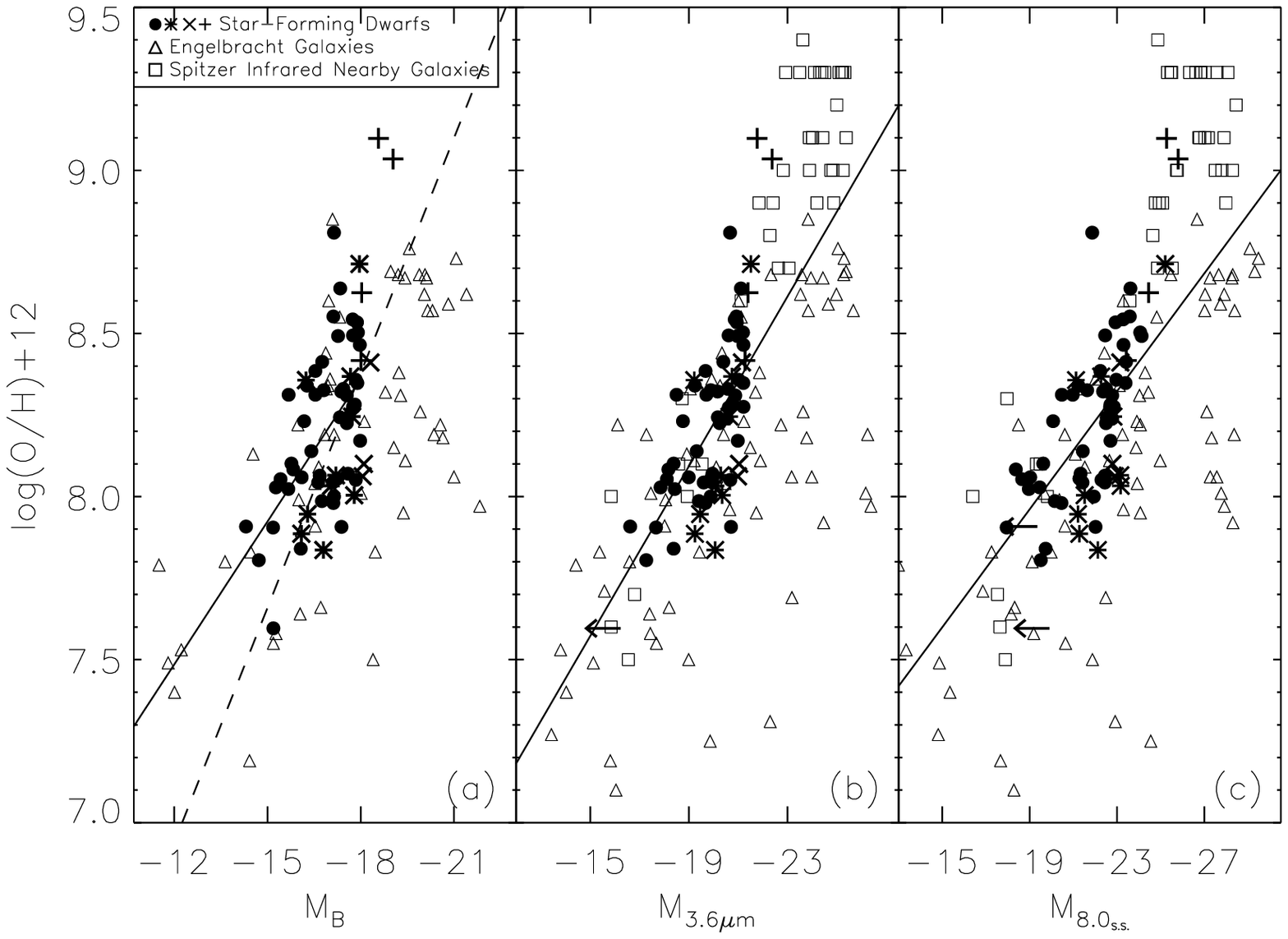}
\caption{Metallicity as a function of absolute magnitude in the $B$, 3.6, and 8.0$_{S.S.}$
\um\ bands. The galaxies in our sample are shown by the same symbols as
in Figure \ref{midc4}. SINGS galaxies \citep{kennicutt2003} are shown as
open squares and star-forming galaxies from \citet{engelbracht2005} are
shown as open triangles. The solid lines indicate the linear least-squares
fits to the data in this sample with $M_{B_0} > -18$. The dashed line in
panel (a) shows the slope of the metallicity--luminosity relation from
\citet{melbourne2002} for the full sample of KISS galaxies.}
\label{newlz}
\end{figure}

Figure \ref{newlz} shows the relationship between metallicity and $M_{B_0}$,
$M_{3.6}$ and $M_{8.0_{S.S.}}$. Most of the galaxies in our sample are sub-solar
([log(O/H) + 12]$_{\odot}$ $=$ 8.69). The luminosity-metallicity relationships
for galaxies fainter than $M_B = -18$ in our sample are determined from a linear
least-squares fit to the data points giving the relations:
\begin{center}
log(O/H) + 12 = $-$0.15($\pm$0.03)$M_{B_0}$ + 5.73( $\pm$0.50) \\
log(O/H) + 12 = $-$0.13($\pm$0.02)$M_{3.6}$ + 5.62( $\pm$0.43) \\
log(O/H) + 12 = $-$0.09($\pm$0.02)$M_{8.0_{S.S.}}$ + 6.24( $\pm$0.34) \\
\end{center}
giving an rms scatter around the fits of 0.20 at $B$-band, 0.18 at 3.6 \um, and
0.18 at 8.0 \um. The slope of these relations is consistent with the
slope determined from 19 data points in Paper I, although we now get a slightly
shallower slope at $B$-band and 3.6 \um\ and a slightly steeper slope at 8.0 \um\
which brings all of the values in all three bands closer together. For
comparison with our $B$-band data, we show the linear least-squares fit 
to the full KISS sample \citetext{including $M_B < -18$,
\citealt{melbourne2002}}. This $B$-band slope is significantly shallower than that
determined for the full KISS sample \citep[$-$0.24]{melbourne2002} and for
Blue Compact Dwarf Galaxies \citep[$-$0.24]{zhao2010}.
This flattening at the low-luminosity end is consistent with other values obtained
for dwarf galaxies \citep{vanzee2006,shi2005,lee2004,richer1995,skillman1989} and
has been previously noted by \citet{melbourne2002}. The flattening of the slope of
this relation at lower luminosity implies that the dwarf galaxies have a relatively
larger metallicity than the more luminous systems. This difference could be driven
by a floor on the metallicity or differences in the astrophysical processes (e.g.,
outflows versus closed-box enrichment) that govern this relationship.

The 3.6 \um\ luminosity--metallicity relationship is expected to be tighter
than the $B$-band relationship because it is less influenced by extinction and it
is a better measure of the stellar mass of the system (as noted in the previous
section, it does not appear to be strongly influenced by PAH emission at 3.3
\um). However, we find only a modest decrease in the scatter between $B$-band and
3.6 \um. Even more surprising, the scatter in the 8.0 \um\ relationship is the
same as for the 3.6 \um\ relationship. We would expect the dependence of the 8.0
\um\ luminosity on the combination of the PAH and dust continuum emission which
are influenced by the SFR, the radiation field, and possibly the
age of the stellar population to increase the scatter. 

\begin{figure}
\epsscale{0.9}
\plotone{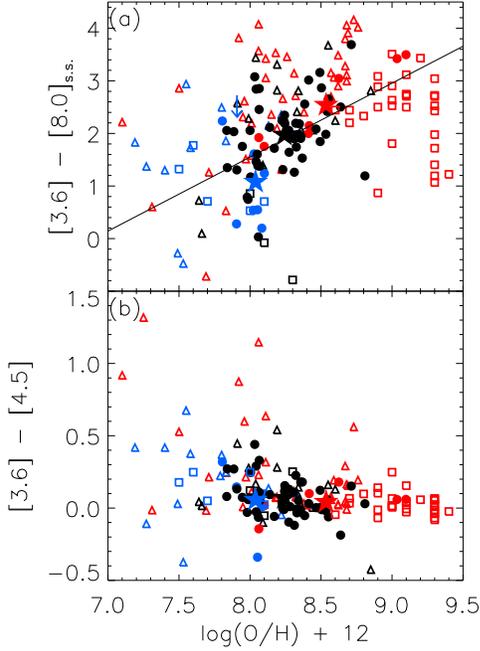}
\caption{[3.6] $-$ [8.0]$_{S.S.}$ and [3.6] $-$ [4.5] color as a function of metallicity. The
galaxies in our sample are shown as filled circles. SINGS galaxies
are shown as open squares and star-forming galaxies from \citet{engelbracht2005}
are shown as open triangles. We divide our sample into three groups based on their 
magnitude at $B$ band. Blue points are low-luminosity sources (\lowmb), black
points are moderate-luminosity sources (\midmb), and red 
points are high-luminosity sources (\mblarge). Filled stars indicate the 
average value for each of these magnitude groups. The solid line in panel (a)
indicates the linear least squares fit to the data with $M_{B_0} > -18$.}
\label{compu3}
\end{figure}

Figure \ref{compu3} shows the relationship between the mid-infrared colors
of galaxies and metallicity for galaxies in our sample, star-forming dwarf
galaxies from \citet{engelbracht2005}, and galaxies from SINGS \citep{kennicutt2003}.
The mass-normalized dust content of
galaxies in Figure \ref{compu3}(a) is correlated with the metallicity, although there is a
lot of scatter in the relationship. Like the color--luminosity relationship, this
correlation was not evident with the smaller sample in Paper I. The linear least-squares fit
on the log scale to the relationship is
\begin{center}
$[3.6] - [8.0]_{S.S.}$ = 1.41($\pm$0.39) $\times$ [log(O/H)+12] $-$ 9.70($\pm$3.23) \\
\end{center}
with an rms scatter around the fit of 0.67 and correlation coefficient 
0.43. The large scatter in this relationship indicates that variations in color
are being driven by more than the effect of metallicity on the dust emission from
the galaxies. The average [3.6] $-$ [4.5] color is 0.1 for the KISS dwarfs. This
color is slightly redder than expected for a stellar SED (see Figure
\ref{compu}) indicating that the color of some of the galaxies is being
influenced by hot dust in the 4.5 \um\ band. As indicated by Figure \ref{compu},
there are galaxies with substantially redder colors than the average indicating
hot dust emission at 4.5 \um, than the average --
primarily luminous systems in the \citet{engelbracht2005} sample. 

\subsection{Effect of Star Formation Rate on the Properties of Star-forming
Dwarf Galaxies} \label{bozomath}

The galaxies in this sample have been selected because they have H$\alpha$
emission detected in the KISS objective-prism survey. In addition, we have made a
luminosity cut to select dwarf systems. While these two cuts limit the
properties of the sample, they still make up a fairly diverse group ranging from
strongly star-forming blue compact dwarf galaxies to more diffuse star-forming
irregular galaxies. Most of the galaxies exhibit SFRs
ranging between 0.001 and 10 $M$\solar\ yr$^{-1}$. For three sources, the SFRs are even lower
than this limit and KISSR 207 has a huge SFR at 85 $M$\solar\ yr$^{-1}$,
although it does not quite make our luminosity cutoff with $M_{B_0}$ = $-$18.06. 

\begin{figure}
\epsscale{1.0}
\plotone{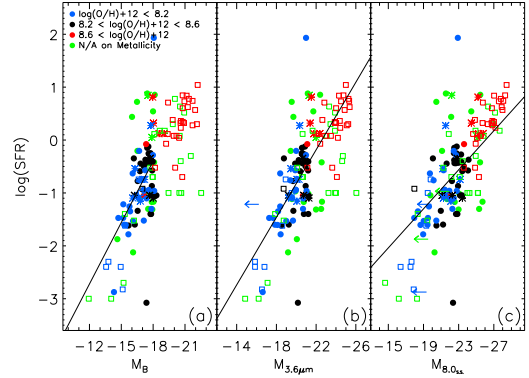}
\caption{SFR, computed from the galaxies H$\alpha$ luminosity,
as a function of absolute magnitude in the $B$, 3.6, and 8.0$_{S.S.}$ $\mu$m bands.
The galaxies in our sample are shown as filled circles (velocity $<$ 9000 \kms) and
asterisks (velocity $>$ 9000 \kms). SINGS galaxies are shown as open squares. We divide
our sample into  three groups based on their metallicity. Blue points are low-metallicity
systems (log(O/H)+12 $<$ 8.2), black points are moderate-metallicity
(8.2 $<$ log(O/H)+12 $<$ 8.6), and red points are high-metallicity (log(O/H)+12 $>$ 8.6).
Green points are galaxies for which metallicity information is not available.
The lines show linear least squares fits to the galaxies with $M_{B_0} > -18$.}
\label{newsf}
\end{figure}

As expected, the SFR in these galaxies is correlated with the blue
luminosity as shown in Figure \ref{newsf}(a). The SINGS galaxies are
included in the plot for comparison over a wider range in luminosity and
SFR. We also show the relationship between the SFR and $M_{3.6}$ and M$_{8.0_{S.S.}}$.
With only two exceptions, all of the high SFR galaxies have high, or unknown
metallicities. However, the middle and low-metallicity sources are well mixed at
the moderate and low SFRs. The relationships between the SFR and luminosity for
our sample are
\begin{center}
log(SFR) = $-$0.39($\pm$0.08)$M_{B_0}$ $-$ 7.37($\pm$1.42) \\
log(SFR) = $-$0.32($\pm$0.07)$M_{3.6}$ $-$ 7.21($\pm$1.36) \\
log(SFR) = $-$0.18($\pm$0.05)$M_{8.0_{S.S.}}$ $-$ 4.81($\pm$1.07) \\
\end{center}
with an rms scatter of 0.66 at $B$-band, 0.66 at 3.6 \um, and 0.65 at 8.0 \um.
There are
two outliers in this plot -- KISSR 207 is a low metallicity galaxy (log(O/H) +
12 = 8.06) with the highest SFR, 85 M\solar\ yr$^{-1}$ in the sample, although
its luminosity, M$_{B_0} =18.06$, is just above our nominal limit. KISSR 1048
is a moderate metallicity source (log(O/H) + 12 = 8.32) with the lowest SFR in
the sample, but it is moderately bright in all bands (M$_{B_0} = -17.4$). 

\begin{figure}
\plotone{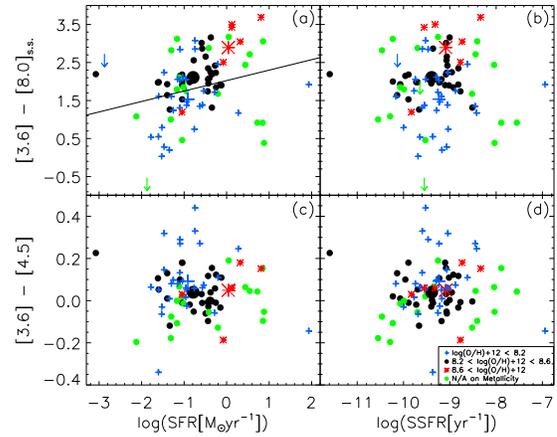}
\caption{[3.6] $-$ [8.0]$_{S.S.}$ and [3.6] $-$ [4.5] as a function of the SFR and the SSFR.
The sample is divided into three groups based on their metallicity. Blue plus signs are
low-metallicity systems (log(O/H)+12 $<$ 8.2), black circles are moderate-metallicity
(8.2 $<$ log(O/H)+12 $<$ 8.6), and red asterisks are high-metallicity (log(O/H)+12
$>$ 8.6). Green circles are galaxies for which metallicity information is not
available. Large plus signs, asterisks, and black filled circles indicate the average
value for each of these metallicity groups. The solid line in panel (a) indicates
the linear least-squares fit to the data with $M_{B_0} > -18$. KISSR 207, the highest SFR source
is brighter than $M_{B_0} = -18$ and is not included in the fit.}
\label{sfrmir2}
\end{figure}

Figure \ref{sfrmir2} shows the relationship between mid-infrared colors and both
the SFR and the specific star formation rate (SSFR; SFR/$M_{stellar}$).
The stellar mass used to calculate the SSFR was
derived from the $B - V$ color and the $B$-band luminosity \citep{bell2001}. There
is a correlation between the [3.6] $-$ [8.0] color and the SFR driven mostly by the
highest metallicity sources in the sample. The linear least squares fit on the log
scale to these data is given by:
\begin{center}
$[3.6] - [8.0]_{S.S.}$ = 0.27($\pm$0.13) $\times$ log(SFR) + 2.02($\pm$0.13) \\
\end{center}
which has an rms scatter of 0.75 and correlation coefficient 0.25. The major
outliers in this plot include the highest SFR source KISSR 207 (blue cross on
the far right) which is more luminous than our nominal cutoff of $M_B$ = $-$18,
KISSR 1048, the lowest SFR source in the sample, and KISSR 238, 541, 2407 which
have high SFRs and bluer [3.6] $-$ [8.0]$_{S.S.}$ colors. The relationship for
the 19 galaxies in Paper I had an rms scatter of 0.56 and correlation coefficient
0.66, so the increase in sample size has increased the number of outliers and,
overall, decreased the correlation.

The relationship between [3.6] $-$ [8.0]$_{S.S.}$ color and SSFR shows virtually
no correlation (correlation coefficient = 0.02). This lack of a correlation is
somewhat surprising given the correlation between 8.0 \um\ luminosity and
SFR since this relation is the same one normalized by stellar mass since 3.6
\um\ luminosity is a good probe of stellar mass. The SSFR in Figure
\ref{sfrmir2} is computed using stellar mass calculated with the \citet{bell2001}
calibration for optical colors, but the plot is almost identical if the stellar
mass is computed using the 3.6 \um\ luminosity with the calibration of
\citet{zhu2010}. SFR is correlated with both 3.6 and 8.0
\um\ luminosities as shown in Figures \ref{newsf}(b) and (c) so the lack of correlation
between [3.6] $-$ [8.0]$_{S.S.}$ color and SSFR in Figure \ref{sfrmir2}
indicates that on a global scale in these galaxies, the higher 8.0 \um\
luminosity in higher SFR galaxies is more strongly driven by the mass of the
galaxy than by the star formation processes (i.e., more massive galaxies are
also the more luminous and higher SFR galaxies in general). This result is also
reflection of the significant contribution made by the dust continuum emission,
which is not a function of the SFR, to the 8.0 \um\ luminosity in these
galaxies. Because this continuum contribution is significant the [3.6] $-$
[8.0]$_{S.S.}$ color is not correlated with SSFR.

\subsection{8.0 $\mu$m Luminosity Function} \label{bozomath}

The 8.0 $\mu$m luminosity function of star-forming dwarf galaxies is a measure
of the contribution of these systems to the total 8.0 $\mu$m luminosity density
in the local universe. We compute the luminosity function based on the 57
galaxies in the sample with velocities between 1000 and 9000 \kms. We exclude the
lowest velocity, and hence most nearby sources because they are detected in such
a small effective volume that they are strongly influenced by the local galaxy
density and they contribute disproportionately to the luminosity function. We
also exclude galaxies with $v_{hel} >$ 9000 \kms\ because they are above our
velocity cut in the 30\dgr\ strip. The
calculation was carried out using the 1/$V_{max}$ method \citep{schmidt1968},
where $V_{max}$ is the volume corresponding to the maximum redshift at which the
galaxy's H$\alpha$ emission line could have been detected by the survey. In
addition, we compute a luminosity function based on an estimate of the stellar
subtracted 8.0 \um\ emission which represents the contribution of PAHs and the
hot dust continuum to the 8.0 \um\ band. The stellar subtracted 8.0 \um\
luminosity is the 8.0 $\mu$m luminosity minus the 3.6 $\mu$m luminosity scaled
down by a factor of 0.227 \citep{pahre2004a}, the stellar contribution from an M0
III star.

\begin{figure}
\plotone{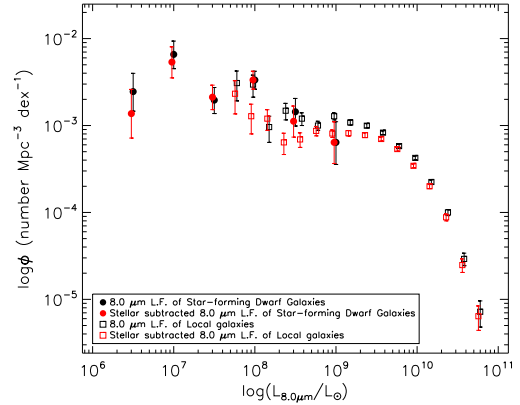}
\caption{Luminosity function for our sources (filled circles) and for a complete 8.0
\um\ selected sample in the Bo\"{o}tes field (\citealp{huang2007}, open squares).
The black points are the full 8.0 \um\ luminosity function while the red points
show the stellar continuum subtracted values. These two sets of points have been
artificially offset (stellar subtracted numbers to left) so that they are both visible
since in most cases they lie on top of one another.}
\label{lf}
\end{figure}

In Figure \ref{lf}, we plot the 8.0 $\mu$m luminosity function for these
galaxies and for an 8.0 \um\ - selected sample in the Bo\"{o}tes field
\citep{huang2007}. The luminosity function error bars for the star-forming dwarf
galaxies are based on Poisson statistics \citep{gehrels1986} which
correspond to 1$\sigma$ Gaussian errors. Because we
restrict our sample to the faint end of the luminosity function, we do not have
a good constraint on the luminosity function's shape. However, the dwarf
galaxy luminosity follows the shape of the full sample within the errors and the
luminosity density (3.1 $\times$ 10$^{6}$ $L_\sun$ Mpc$^{-3}$) is comparable
to the \citet{huang2007} luminosity (2.9 $\times$ 10$^{6}$ $L_\sun$ Mpc$^{-3}$)
density below 10$^{9}$ $L$\solar\ indicating that 8.0 \um\ and H$\alpha$
selection recover the same 8.0 \um\ luminosity density, when their respective
selection limits are taken into account. 

The fraction of the total 8.0 \um\
luminosity density that comes from star-forming dwarf galaxies is derived from
the integration of the luminosity function. We find the luminosity density to be
3.1 $\times$ 10$^{6}$ $L_\sun$ Mpc$^{-3}$ or 2.9 $\times$ 10$^{6}$ $L_\sun$
Mpc$^{-3}$ if we use the stellar continuum subtracted values. This number is in
contrast to the luminosity density derived for all galaxies in the Bo\"{o}tes
field of 2.8 $\times$ 10$^{7}$ $L_\sun$ Mpc$^{-3}$ or 2.3 $\times$ 10$^{7}$
$L_\sun$ Mpc$^{-3}$ if stellar continuum subtracted values are considered
\citep{huang2007}. The relationship between these SFRD
(SFRD) values indicates that star-forming dwarf galaxies contribute $\sim$12\%
of the 8.0 $\mu$m luminosity in the local universe which is comparable to the
fraction of the H$\alpha$ luminosity density that they contribute. 

\section{SUMMARY} \label{bozomath}

We have presented {\it Spitzer}/IRAC observations of 86 star-forming dwarf galaxies
selected from the KISS optical objective prism survey. We build on the previous
study of a much smaller number (19) of these galaxies (Paper I)
to increase our understanding of the infrared emission from low-luminosity,
low-metallicity galaxies. In addition we make use of the statistically complete
nature of this sample to examine the luminosity function and luminosity density
of these galaxies at 8.0 \um. 

Overall the dust emission in these galaxies is clearly affected by both the
metallicity and the SFR. We see much better correlations with the 8.0 \um\
luminosity than we do with the [3.6] $-$ [8.0] color, especially in the case of
the SFR. Nevertheless, there are correlations between the [3.6] $-$ [8.0] color
and both luminosity and metallicity (and to a much lesser extent SFR) than was
evident from the sample of 19 galaxies. More specifically we find the following.

\begin{itemize}

\item The median 8.0 \um\ luminosity in the sample is 0.004\lstar\ while
the median value at $B$-band and 3.6 \um\ is 0.06\lstar. This
lower luminosity with respect to \lstar\ at 8.0 \um\ implies that there is
a deficit of hot dust and PAH emission in these galaxies with respect to the
majority of galaxies.

\item As we found previously with the smaller sample, star-forming dwarf
galaxies have much bluer than average optical colors and span a wide range of
infrared colors. In [3.6] $-$ [8.0], the galaxies range from very red objects to
the colors of early type galaxies. The fraction of very red star-forming dwarf galaxies
is almost 4 times higher than it is for SDWFS galaxies with similar optical colors.

\item The larger sample investigated in this paper indicates that the [3.6] $-$
[8.0] color is correlated with luminosity, metallicity, and to a much lesser
extent SFR. These relationships were not evident with the smaller sample
presented in Paper I. 

\item The luminosity--metallicity relationship for star-forming dwarf galaxies
has a shallower slope than what is found for the full KISS sample, consistent
with the low-luminosity slope of the SDSS relationship. The change in the slope
of this relation with luminosity may indicate either a floor in
the metallicity of galaxies or differences in the astrophysical
processes (e.g., outflows versus closed-box enrichment) that govern this
relationship.  

\item The relationship between luminosity, particularly at 8.0 \um, and SFR
indicates that the 8.0 \um\ luminosity is another probe of the SFR in
galaxies despite the fact
that the [3.6] $-$ [8.0] color shows a much stronger correlation with
metallicity than with SFR. While there is a small correlation between the [3.6] $-$
[8.0] color and the SFR, there is no apparent correlation when we consider the
relationship between color and specific SFR.

\item The consistency between the 8.0 \um\ luminosity function and density
derived indicates that below
10$^{9}$ $L$\solar, nearly all the 8.0 \um\ luminosity density comes from dwarf
galaxies that exhibit strong H$\alpha$ emission -- i.e., 8.0 \um\ and \ha\
selection identify similar galaxy populations. These galaxies span a similar
range of SFRs and have similar space densities.

\item The 8.0 \um\ luminosity is, on average, a good indicator of SFR
in these dwarf systems as measured by the comparison of the SFRD
computed from the  8.0 \um\ and H$\alpha$ emission.
This is despite the deficit of hot dust and PAH emission in the 
star-forming dwarf galaxies (see the first point), the poor correlation between
the [3.6] $-$ [8.0] color and the SFR, and the large spread in the SFR versus 8.0
\um\ luminosity relationship.  

\end{itemize}

\acknowledgments
This work is based in part on observations made with the {\it Spitzer
Space Telescope}, which is operated by the Jet Propulsion Laboratory,
California Institute of Technology under NASA contract 1407.
Support for this work was provided by NASA through an IRAC GTO award
issued by JPL/Caltech under contract 1256790.
J. J. S. gratefully acknowledges support for the KISS project from the NSF
through grants AST 95-53020, AST 00-71114, and AST 03-07766.
This work also made of data products provided by the SDSS.
The SDSS is managed by the Astrophysical Research
Consortium for the Participating Institutions. The Participating Institutions
are the American Museum of Natural History, Astrophysical Institute
Potsdam, University of Basel, University of Cambridge, Case Western
Reserve University, University of Chicago, Drexel University, Fermilab,
the Institute for Advanced Study, the Japan Participation Group, Johns
Hopkins University, the Joint Institute for Nuclear Astrophysics, the Kavli
Institute for Particle Astrophysics and Cosmology, the Korean Scientist
Group, the Chinese Academy of Sciences (LAMOST), Los Alamos National
Laboratory, the Max-Planck-Institute for Astronomy (MPIA),
the Max-Planck-Institute for Astrophysics (MPA), New Mexico State
University, Ohio State University, University of Pittsburgh, University of
Portsmouth, Princeton University, the United States Naval Observatory,
and the University of Washington.

\end{document}